\documentstyle[sprocl,epsfig]{article}

\begin{document}
\title{BEYOND THE PARTON CASCADE MODEL:\\
Klaus Kinder-Geiger and VNI}
\author{BERNDT M\"ULLER \\
	Department of Physics \\
	Duke University \\
	Durham, NC 27708-0305, USA}

\maketitle 

\abstracts{
I review Klaus Kinder-Geiger's contributions to the physics of relativistic
heavy ion collisions, in particular, the Parton Cascade Model. Klaus
developed this model in order to provide a QCD--based description of 
nucleus--nucleus reactions at high energies such as they will soon
become available at the Brookhaven Relativistic Heavy Ion Collider.  
The PCM describes the collision dynamics within the early and dense 
phase of the reaction in terms of the relativistic, probabilistic 
transport of perturbative excitations (partons) of the QCD vacuum.
I will present an overview of the current state of the numerical 
implementations of this model, as well as its predictions for nuclear 
collisions at RHIC and LHC. }

\section{Introduction}

Klaus Kinder-Geiger (KKG), who perished in the crash of Swissair flight 
111 near Halifax on September 2, 1998, was a brilliant theoretical
physicist and one of my dearest friends. He was one of those few
human beings who are truly irreplaceable, not because no one else
could carry on Klaus' research, but because of the unique way he did 
physics and almost everything else in life. Klaus combined the abstract 
mind of the scientist who works on deep and esoteric questions of 
nature with the wild mind of the artist who is driven to create and
perform in extraordinary ways. Klaus became famous as a physicist
for the work he did on the parton cascade model of relativistic
nuclear reactions.
But he was equally famous as an extravagantly creative and perceptive
person in the communities where he lived and made friends. I always
thought that knowing Klaus was the closest I would ever get to meeting
in person one of the French existentialists whose novels I had read
as a student.

\begin{figure}[htb]
\vfill
\centerline{
\begin{minipage}[t]{.57\linewidth}\centering
\mbox{\epsfig{file=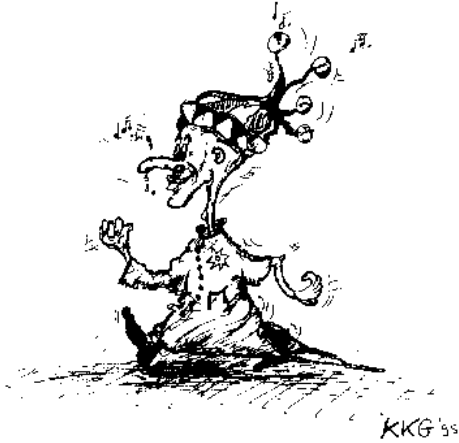,width=.95\linewidth}}
\end{minipage}
\hspace{.04\linewidth}
\begin{minipage}[b]{.37\linewidth}\centering
\caption{The ``Wunderbar World of KKG'' sketch greeting visitors to
Klaus' World Wide Web site at BNL.}
\label{Muller:fig1}
\end{minipage}}
\end{figure}

\begin{figure}[htb]
\centerline{\mbox{\epsfig{file=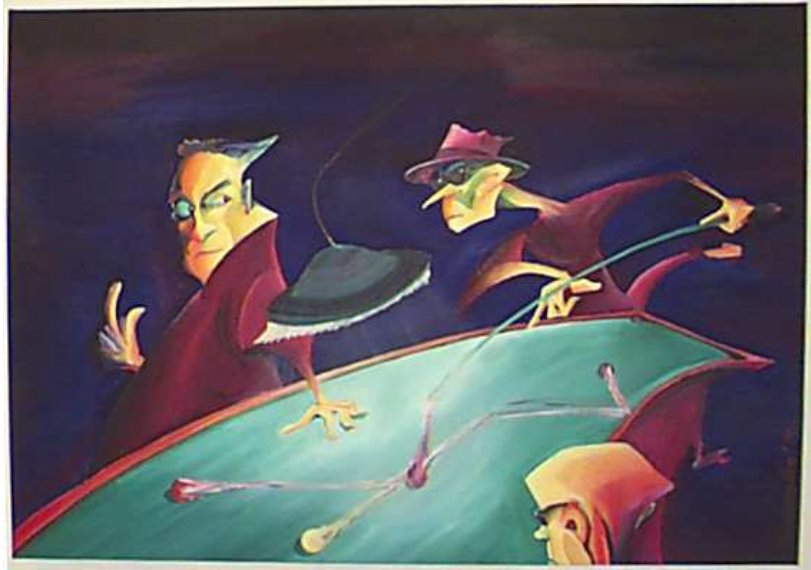,width=.9\linewidth}}}
\caption{The Pool Players. By {\sl Vincent De Cucurullo}, 1986}
\label{Muller:fig2}
\end{figure}

As many of you know, Klaus very much liked to draw sketches of himself
and the world around him. The one entitled ``The Wunderbar World of 
KKG'', featured prominently in his home page on the World Wide Web 
(Figure \ref{Muller:fig1}).  It captures nicely how Klaus viewed 
himself: always inspired to explore unfamiliar territory, uncover 
new ideas, and gather new experiences. 
Klaus believed that what really counted in life was the unusual, and he
dared to live by his belief. When he was still a young student at
the University of Frankfurt, before I knew him, Klaus had
tried his hand at painting. His paintings are highly expressive and
almost haunting; having seen them once one is not likely to ever 
forget them. One of his paintings, shown in 
Figure \ref{Muller:fig2}, is particularly
remarkable: it anticipates Klaus' greatest contribution to physics.
Clearly, the subject of the painting must be a parton cascade, as
Klaus imagined it in his artistic mind in 1986.\footnote{Of course,
no one - not even Klaus himself - would have had any notion of this
at the time the scene was painted. But the artist sometimes envisions
novel ideas and concepts long before the scientist. What is unusual
here is that the artist became the scientist!} 

The hustler, I am convinced, is depicting Klaus himself. 
(It is anyone's guess whom the two other figures represent. 
Some similarities with well known members of the heavy ion physics 
community are unmistakable.) Lest anyone doubts this
interpretation of the painting, let me point out that it is signed
``Cucurullo '86''. Vincent de Cucurullo, short: Vinnie, was Klaus' 
artistic pseudonym by which he signed his paintings. The name of his 
celebrated parton cascade model code, VNI, is a rendering of the 
artist's name under the cloak of a scientific acronym.

\section{The Parton Cascade Model} 

The parton cascade model (PCM) was proposed by Klaus and me \cite{Muller:GM92} 
in 1990-91 and developed much further during the following years
by Klaus in a remarkable series of
papers.\cite{Muller:Gei92a,Muller:Gei92b,Muller:Gei94,Muller:Gei95}
Our aim was to describe the energy deposition, thermalization, and chemical 
equilibration of matter in high energy nuclear collisions, and to provide 
a full space-time picture of the collision up to the moment when individual hadrons 
are formed.  The model was, at least originally, not conceived as an ``event 
generator'' that would predict a full set of hadron momentum distributions in 
the final state.  

In order to enable experimental predictions, Klaus decided to 
develop the parton cascade model code VNI, which contains the implementation 
of a hadronization scheme in the framework of a parton coalenscence model.  
This required a number of compromises and ran somewhat counter to the original 
purpose of the PCM, namely, to explore the range of validity of perturbative 
QCD in nuclear reactions.  Nevertheless, the predictions of the PCM with an 
added hadronization stage have been, and continue to be, very useful. We simply
do not know how to do better at the present time.

The conceptual basis of the PCM is the inside-outside cascade model 
\cite{Muller:AKM80} of high-energy hadron reactions, which implements the concept 
that new matter produced in hadronic interactions at high energy is formed outside 
the intersecting world-tubes of the colliding hadrons.  Bjorken's hydrodynamical 
model \cite{Muller:Bj83} was devised to describe the evolution of this newly formed 
matter in the central space-time region after thermal equilibration.  
Beginning in the mid-1980's it was realized that the deposition of energy into 
this region may be, at least partially, described in terms of concepts based 
on perturbative QCD (minijets) when the energy of colliding heavy nuclei 
becomes very large.\cite{Muller:HK86,Muller:BM87,Muller:KLL87,Muller:EKL}  
A computer code (HIJING) incorporating some of these ideas was developed by 
Gyulassy and Wang.\cite{Muller:GW,Muller:Wang97}

The parton cascade model combined these ideas into one unified scheme for 
the description of the space-time evolution of matter in nuclear reactions.  
Its three main ingredients are: 

\begin{enumerate} 
\item  
The {\em initial state} is viewed as incoherent ensemble of partons determined 
by the nuclear parton distribution functions $q_f(x, Q^2)$ and $g(x, Q^2)$, 
where the subscript {\it f} denotes the quark flavor and 
$g(x,Q^2)$ stands for the gluon distribution.  $x = p_z/P$ is the 
longitudinal momentum fraction of the nucleon carried by the parton, and $Q^2$ 
is the parton ``scale'' or virtuality.  Before any interaction occurs, $Q^2$ 
is generally taken as space-like.  Our knowledge about the space-time 
structure of the nuclei before the collision and our limited information about 
the intrinsic transverse momenta of partons is then used to construct a model 
for the six--dimensional phase space distributions of partons before the 
interaction: $q_f (r, p), g(r, p)$.  
The parton distributions are conveniently initialized 
at the  scale $Q^2_0 = \langle p^2_T \rangle_{\rm coll}$ of the average 
momentum scale of the primary parton-parton interactions. 

\item  
The {\em time evolution} of the parton phase distributions is governed by a 
relativistic Boltzmann equation with ``leading-log'' improved lowest-order 
collision terms.  Only binary interactions are allowed, but the final state 
can have (and generally has) more than two particles.  As is well known, the 
higher-order improvement of the cross sections by means of the leading 
logarithmic approximation is equivalent to the scale evolution of the parton 
distributions according to the DGLAP equation.  Motivated by quantum 
mechanical considerations, the space-time picture of parton propagation 
before and after interactions is closely related to their off-shell 
propagation: the formation of a parton with virtuality scale $Q$ takes a 
time $\tau_f(Q) \approx \hbar/Q$. 

\item 
When the parton distributions become sufficiently dilute, they {\em hadronize}.
In VNI the hadronization is described by a clustering algorithm, followed by 
the decay of excited hadrons.  The transition is assumed to occur when the 
average virtuality of the partons falls below a critical value $Q_{\rm crit}
\approx 1$ GeV, because partons no longer scatter with sufficient energy
to allow for the collisions to be described by perturbative QCD.
\end{enumerate}

From a gradient expansion of the evolution equation for the parton Wigner 
distribution two equations can be derived.\cite{Muller:Gei96}  
The first equation 
\begin{equation} 
p^\mu \frac{\partial}{\partial r^\mu} F_i(r, p) = C_i (r, p)
\label{Muller:eq.star} 
\end{equation}
describes the free propagation of partons which is intermittently modified by 
interactions given by the binary collision terms $C$.  The second equation 
\begin{equation}
p^2 \frac{\partial}{\partial p^2} F_i (r, p) = S_i (r, p)
\label{Muller:eq.twinkle} 
\end{equation} 
describes the evolution of the parton distributions with respect to virtuality 
or ``off-shellness'' $p^2$.  This equation is a generalization of the usual 
mass-shell condition $F(r, p) \sim \delta (p^2-m^2)$ to the case where the 
on--shell particle distribution cannot be defined.  $S_i (r, p)$ describes 
the splitting of single off-shell partons into two partons of smaller 
virtuality. 

\begin{figure}[htb]
\vfill
\centerline{
\begin{minipage}[t]{.47\linewidth}\centering
\mbox{\epsfig{file=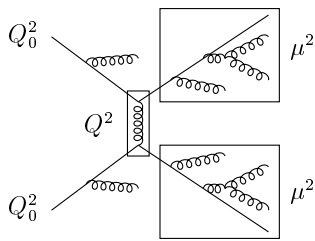,width=.9\linewidth}}
\end{minipage}
\hspace{.04\linewidth}
\begin{minipage}[b]{.47\linewidth}\centering
\caption{Graphical representation of the QCD transport equations (1,2) 
defining the parton cascade model.  The $2 \to 2$ scattering process at 
momentum scale $Q^2$ is followed by the virtuality evolution from $Q^2$ 
to $\mu^2$.}
\label{Muller:fig3}
\end{minipage}}
\end{figure}

The two equations can be viewed as quantitative representation of Feynman 
diagrams of the type shown in Fig.~\ref{Muller:fig3}.  
The collision term $C_i$ is represented by the binary collision diagram 
contained in the box at the center of the complex Feynman diagram, whereas 
the splitting terms $S_i$ are represented by the branchings of the initial-- 
and final-state partons.\footnote{Strictly speaking, $S_i$ describes the 
differential branching probability for an infinitesimal change in virtuality; 
the diagram is an integral representation of $S_i$.}  The Feynman diagram 
of Fig.~\ref{Muller:fig3} is finite only if the virtualities of all 
final--state partons are limited from below by 
some infrared cut-off $\mu^2$.  In an isolated event $\mu^2$ describes the 
{\em hadronization scale}, i.e. the virtuality scale below which partons can 
no longer be considered as approximately free, perturbation quanta.  In a 
dense medium, where partons rescatter often, $\mu^2$ is determined by the 
frequency of rescatterings (see Section 3.2). 

In the leading-logarithmic approximation (LLA), the differential cross 
section described by the Feynman diagram of Fig.~\ref{Muller:fig3} 
factorizes into a product of terms 
for each of the two incoming and outgoing branch processes and one for the 
binary scattering process.  Each branching term, in turn, is represented by a 
product of factors describing the individual branching events and the 
probabilities for the partons {\em not} to branch further in between.  In 
other words, the Feynman diagram of the type shown in Fig.~\ref{Muller:fig3} 
defines a Markov process and, hence, can be described by a 
probabilistic one--body transport equation.  This statement is no longer 
valid, if one tries to go beyond the LLA.  However, certain effects beyond 
the LLA can still be described in terms of conditional probabilities, such as 
angular or $k_T$--ordering of gluons and certain soft--gluon interference 
effects.  These effects are quantitatively important and have been 
incorporated into parton cascade codes.\cite{Muller:Gei92b}  

It is important to note that 
parton splittings are related by unitarity to loop diagrams that describe 
the running of the strong coupling constant $\alpha_s (Q^2)$.
Both splittings and $\alpha_s$--running are described consistently in the LLA, 
which therefore satisfies the unitarity condition.  In plain terms, the 
combined probability for all $2\to n$ parton diagrams with $n \geq 3$ reduces 
the probability for the occurrence of a $2 \to 2$ scattering and so on. 
By summing all $2 \to n$ free diagrams, but not including the associated loop 
diagrams, unitarity would be violated.  This violation
leads to the divergence of the sum over $n$ already at rather small 
parton center-of-mass energies, even in the presence of an infrared cut-off 
for the internal propagators.\cite{Muller:XS94}

In the following sections I will discuss two important issues:  
\begin{enumerate} 
\item 
The space--time picture governing the initial--state parton distributions.  
This issue is closely connected with the problem of the decoherence of the 
initial parton wavefunctions. 
\item 
The problem of infrared divergences of the perturbative parton cross 
sections.  This issue is closely related to in--medium corrections of these 
cross sections, as well as to coherence properties of the initial--state 
wave functions.
\end{enumerate} 
The approach to local thermal equilibrium has been extensively studied within 
the framework of the parton cascade 
picture.\cite{Muller:Wang97,Muller:EW94,Muller:HW96} Without 
repeating the detailed arguments here, let me just state that the PCM approach 
predicts a very short kinetic equilibration time, $\tau_{\rm th} \ll 1$ fm/c, 
which is confirmed by full numerical calculations.\cite{Muller:Gei92a}

\section{Initial--state space--time picture} 

The probabilistic interpretation of parton distributions measured in 
deep--inelastic scattering is based on a summation over all final hadronic 
states.  A similar interpretation of one--body distributions arising in 
transport theory is grounded on the low--order truncation of the (BBGKY) 
hierarchy of Green functions and on an expansion in powers of $\hbar$.  The 
validity of this picture ultimately relies on the separation of time scales 
in a dynamic process.  Although these issues are generally well 
known,\cite{Muller:MR96}, 
their implications for nuclear parton cascades have not been fully explored.  
Recent advances \cite{Muller:Baier} in our understanding 
of multiple scattering in QCD have shed some light on the intricacies of 
the formation time concept in non--abelian gauge theories, but it needs 
to be better understood how these results can be consistently incorporated
into a probabilistic transport theory.

The original parton cascade model relied on some basic assumptions about 
initial parton distributions in space--time.\cite{Muller:EW94}  Denoting the parton 
light--cone momentum by $p^+$, the parton distributions were assumed to be 
distributed longitudinally according to the uncertainty relation: 
$\Delta p^+ \Delta x^- \geq \hbar$. 
Soft partons have the widest distributions in the variable $x^-$, 
and are assumed to travel both ahead of and behind the Lorentz 
contracted valence quark distributions.  The argument is that this will not 
violate causality, because soft partons are emitted at a long distance before 
the collision and, travelling at the speed of light, can arrive significantly 
ahead of the quarks that emitted them. 

The space--time picture of soft partons has been put on a much firmer 
foundation in recent years by the work of McLerran, Venugopalan and others on 
the random light--cone source model (RLSM).\cite{Muller:MV94}  In this model, one 
views the valence quarks constituting the fast--moving nucleus as a thin, 
Lorentz contracted sheet of locally random color sources.  The color source 
is locally random, because valence quarks from several nucleons contribute at 
the same point in transverse space.  The area density of color sources is 
given by $\mu^2 = 3g^2A/\pi R^2$, where $A$ is the nucleon number, $R$ the 
nuclear radius, and $g$ the QCD coupling constant.  
Clearly $\mu$ grows as $A^{1/6}$, hence can be considered
as a (potentially) large scale for sufficiently heavy nuclei.\footnote{
In practice, $\mu \leq 1$ GeV even for the heaviest nuclei, even if the
``hard'' component of the gluon distribution is included in the color source.}
$\alpha_s (\mu^2) \ll 1$ can serve as the coupling parameter for a new type 
of perturbative expansion.  Formally, the model maps into the problem of 
weakly coupled QCD in the presence of a random two--dimensional color source. 

As shown by Kovchegov,\cite{Muller:Kov96} this model can be rigorously derived 
with standard light--cone techniques, which permit an explicit representation of 
the Gaussian ensemble of color sources.  This representation can also be used 
to calculate the perturbative emission of soft gluons in collisions between 
two nuclei, described as encounter of two sheet--like clouds of valence 
quarks.\cite{Muller:KR96,Muller:MMR97}  
At leading order this soft gluon radiation is given by:
\begin{equation}
\frac{dN_g}{dy d^2k_ \perp d^2b} 
 = \frac{ 4\alpha^3_s}{\pi^2 k^2_\perp} 
\frac{N^2_c -1}{N_c} \langle T_{AB}(b) \rangle 
 \int d^2g  
\frac{ F(qa) F(|k-g| a)}{ q^2(k-q)^2}
\end{equation}
where $F(qa)$ is the color--dipole form factor of the nucleon and $T_{AB}(b)$ 
denotes the nuclear profile function.  It can be shown \cite{Muller:GML97} that this 
classical gluon radiation matches smoothly onto the perturbative minijet 
production of gluons at higher $k_\perp$. 

Going beyond the classical approximation by including gluon--loop diagrams 
leads to a better and more rigorous understanding of the space--time 
distribution of soft gluons in a heavy nucleus.\cite{Muller:KMW97}  The quantum 
corrections can be formulated in the framework of a space--time analogue of 
the renormalization group equations, describing the cascade of gluon emission 
leading to a power--law enhancement of soft gluons similar to the BFKL 
equation.\cite{Muller:KLW97} 
The RLSM approach also describes screening effects in the parton distribution 
at small $k_\perp$.  The precise origin of this saturation has recently been
elucidated by Kovchegov.\cite{Muller:Kov98}

The picture that emerges is the following: Gluons in the classical field 
generated by the valence quarks are fully Lorentz contracted by the Lorentz 
factor $\gamma$ associated  with the colliding nuclei, but gluons spawned by 
a splitting of those primary gluons experience only a partial Lorentz 
contraction of order $x\gamma$, where $x$ is the momentum fraction carried by 
the parent gluon.  As the branching process evolves to softer and softer 
gluons, the spatial extent of this gluon cloud becomes more and more diffuse 
in the light--cone variable $x^-$.  This result confirms the intuitive 
picture embodied in the original parton cascade model, and provides a 
quantitative formulation of it. 

\section{In--medium effects} 

In--medium effects on parton--parton interactions are essential to the 
viability of the parton cascade model.  The application of perturbative QCD to 
nucleon--nucleon collisions requires the introduction of {\em ad hoc} cut-offs 
describing {\em nonperturbative} QCD effects, such as quark confinement and 
chiral symmetry breaking.  In--medium effects, which grow rapidly in size as 
function of $A$, produce {\em perturbative} cut-offs when the density of the 
medium becomes sufficiently high.  
For example, QCD is known to become perturbative\footnote{Some 
nonperturbative effects remain even at high $T$, precisely 
because static magnetic interactions are not screened by perturbative 
in--medium interactions.} at high temperature when the color--electric screening 
mass $\mu_{\rm D} \gg \Lambda_{\rm QCD}$.  

Although in--medium effects work in favor of the parton cascade model,
they are not easily incorporated in practice.  The problem is that 
in--medium effects are quite complicated and not easily treated correctly.  
The two main in--medium effects that are known to provide effective infrared 
cut-offs to perturbatively divergent parton interactions are: 
\begin{itemize}
\item 
color--electric screening, which suppresses soft $2\to 2$ scattering 
amplitudes; 
\item 
gluon radiation suppression, which reduces $2 \to 3$ (and $2 \to n)$ 
amplitudes with soft gluons in the final state.  
\end{itemize} 
Dynamical screening, at lowest order, is described by the in--medium 
contributions to the one-loop gluon polarization function: 
$$
\centerline{\mbox{\epsfig{file=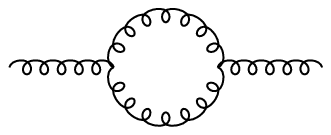}}  }
$$
At moderately high $q_\perp$, the gluon population grows like 
$n(k) \approx (A_1 A_2)^{1/3}$, providing a screening scale $\mu(A)$ 
that increases rapidly with the size of the nuclei participating in
the collision.\cite{Muller:BMW92}

Radiation suppression, also known as the Landau--Pomeranchuk--Migdal (LPM) 
effect, is a much more complicated mechanism.  Its theoretical description 
requires a good understanding of the multiple scattering problem in QCD.  
Considerable progress has recently been made in this area, especially 
through the work of Baier et al.\cite{Muller:Baier}, Zakharov 
\cite{Muller:Zak} and others.\cite{Muller:KHN}  
The main difference between QCD and the well-known case of QED is that 
a radiated gluon also rescatters in the medium at the same order in 
$\alpha_s$ as the radiating particle; this is not so for a photon 
radiated by a fast charge moving in an electromagnetic plasma. 

Revisiting the diagram shown in Fig.~\ref{Muller:fig3}, the parton cascade 
model requires infrared cut-offs for both the central $2 \to 2$ scattering 
matrix element and each of the four branching cascades.  This is where the 
in-medium effects help: at high density one expects that the modifications 
of the elementary scattering amplitude ensure an infrared safe behavior.  
To date, two attempts have been made to practically implement the action 
of these in-medium effects: 
\begin{enumerate} 
\item
In the self--screened parton cascade model (SSPCM \cite{Muller:EMW96}), the 
color--electric screening scale $\mu(p_T)$ was calculated self-consistently 
for primary parton interactions only, and the further evolution of the 
parton plasma was described in the framework of the hydrodynamical model.
\item
In Klaus' PCM code VNI \cite{Muller:VNI}, the space--time picture of 
parton interactions is linked to the virtuality evolution of partons.  
A new interaction is permitted if its momentum transfer exceeds the 
virtuality scale of the participating partons at that time.  Soft
radiation is suppressed in the medium by this rule, because intermediate 
collisions continue to reset the parton virtuality to that of the latest
collision.
\end{enumerate} 
Both of these approaches are based on specific assumptions about the relation
between the space-time and virtuality evolution of off-shell components of 
the parton distributions.  One way to study this issue rigorously is the 
Wigner function representation.  First results \cite{Muller:BD98} obtained by 
this method are interesting, but do not yet fully address the complications 
encountered in a QCD parton cascade.  A more direct approach to the problem 
of the space--time picture of off-shell quantum fluctuation is based on a 
modification of the QCD evolution equations to include an infrared 
scale.\cite{Muller:Gei96,Muller:MS98}   In free space this infrared scale is 
determined by properties of the final state (hadronization scale); in a
medium it is determined by screening effects. 

\subsection{Self--screened parton cascade} 

Here one considers the scattering of an initial state parton as completed 
after a time $\tau(p_T)$ which depends on the momentum transfer in the 
reaction.  The uncertainty relation suggests $\tau(p_T) \sim \hbar /p_T$.  
(We will drop the factor $\hbar$ in the following.)  The scattered partons 
are then assumed to screen the scattering processes that involve a smaller 
momentum transfer:
\begin{equation}
\mu^2_{\rm D}(p_T) 
= \frac{3}{\pi^2} \alpha_s (p^2_T) \int^\infty_{p_T}  d^3k 
|\nabla_k n(k)|.
\end{equation}
The density of partons scattered at $p_T$ is, in turn, influenced by the 
screening because the differential cross section depends on $\mu_0$: 
\begin{equation}
\frac{d\hat{\sigma}}
      { dp^2_T}     \sim 
\frac{\alpha_s (p_T)^2} 
     {(p^2_T + \mu^2_{\rm D}(p_T))^2}  \left|M(\hat{s}, \hat{t})\right|^2.
\end{equation}
If $\mu_{\rm D}(p_T)$ becomes large enough at low $p_T$, so that 
$d\hat{\sigma}/dp^2_T$ remains perturbatively small, the coupled set of 
equations can be integrated down to $p_T = 0$.  Since the rapidity density
of scattered partons grows as $(A_1A_2)^{1/3} \ln \sqrt{s}$, this 
condition requires large $A$ and high energy.  The SSPCM concept
has recently been investigated more formally by Makhlin and
Surdutovich\cite{Muller:MS98} within the framework
of the closed--time--path formalism for real-time Green functions.
In this framework, the final state mass-shell condition for emitted
partons regulates the infrared divergences of single-particle
correlation functions, such as the parton densities. Properly carried
through, this concept leads to a self-consistent equation for the
gluon screening mass (plasmon mass) similar to the one used in the
SSPCM.

Quantitatively, one finds that $\mu_{\rm D}$ approaches about 1 GeV at 
low $p_T$ in Au + Au collisions at RHIC energy (100 GeV/u) and 1.5 GeV 
at LHC energy (2.75 TeV/u).  The differential minijet cross section as 
function of $p_T$ peaks at about the same value, clearly showing the 
improved infrared behavior of the self-screened parton cascade.  The 
total deposited energy within one unit of rapidity and after a 
characteristic formation time of 0.25 fm/c is 
$\epsilon_0 \approx$ 60 GeV/fm$^3$ (RHIC) and 
$\epsilon_0 \approx $ 430 GeV/fm$^3$ (LHC).
The conditions established by the SSPCM can, therefore, be taken as 
initial conditions for the thermal and chemical evolution of a 
quasi-equilibrated parton plasma. 

The kinetic equations for the evolution of such a plasma were 
derived by Bir\'o et al. \cite{Muller:Biro93} and by Xiong and Shuryak
\cite{Muller:XS94}.  
Extensive calculations \cite{Muller:SMM97}, including longitudinal 
and transverse expansion, have shown that the plasma cools down to 
the critical temperature of QCD ($T_c \approx 150$ MeV) after 5 fm/c 
(RHIC) and 10 fm/c (LHC).  The emission of electromagnetic probes by 
such an evolving QCD plasma has also been calculated.\cite{Muller:SMM97}

\subsection{Monte-Carlo space--time cascade} 

\begin{figure}[htb]
\vfill
\centerline{
\begin{minipage}[t]{.47\linewidth}\centering
\mbox{\epsfig{file=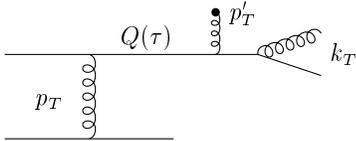,width=.9\linewidth}}  
\end{minipage}
\hspace{.06\linewidth}
\begin{minipage}[b]{.47\linewidth}\centering
\caption{Illustration of in-medium suppression effects incorporated in the 
VNI code.  The virtuality of a scattered parton evolves with time, $Q(\tau)$. 
 Sequential scatterings or branchings are suppressed, if $Q(\tau)$ is too 
large.}
\label{Muller:fig5}
\end{minipage}}
\end{figure}

The statistical implementation of parton cascades in the Monte-Carlo 
simulation code VNI \cite{Muller:VNI} achieves an improved infrared 
behavior through heuristic rules that suppress certain interactions 
on the basis of kinematic considerations.  The first rule asserts that 
independent scattering events involving the same parton require a 
sufficient time separation so that the time between scatterings is 
larger than the duration of the individual events.  With the duration 
of an interaction again defined as $\tau (p_T) \sim p^{-1}_T$, where 
$p_T$ is the momentum exchange, this requires that the time between 
interactions $\Delta \tau > \tau (p_T)$.  Another way of ensuring this 
condition is to endow a parton after a scattering by $p_T$ with an 
initial virtuality $Q_0 = p_T$, which then gradually decreases with 
time as $Q(\tau) = Q_0 \tau(p_T)/\tau$.  A subsequent scattering with 
$p^\prime_T$ requires that $p^\prime_T > Q(\tau)$ at the moment of the 
interaction.  A second similar rule suppresses soft parton splittings
in the presence of multiple 
scattering.  Again, this rule can be formulated in terms of a parton 
virtuality that decreases with time between scatterings and is reset by
each new interaction (see Fig.~\ref{Muller:fig5}).

Although VNI still contains ``arbitrary'' infrared cut-off parameters 
(determined by a comparison with nucleon-nucleon interactions), these are 
needed to limit soft scatterings in the initial set of parton interactions, 
and they become important toward the end of the cascade evolution when the 
parton plasma becomes more and more dilute.  The cut-off parameters 
effectively determine the end of the cascade evolution and, by suppressing 
soft interactions, they trigger the hadronization of parton clusters. 

The fact that this improved parton cascade is much less dependent on 
arbitrary cut-off parameters makes it possible to explore applications 
of this model at lower energies where nucleon--nucleon collisions provide 
little information.  Recently, Geiger and Srivastava \cite{Muller:GS97} have 
studied the predictions that the VNI code makes for nuclear collisions in 
the energy regime of the CERN-SPS.  While most of the particle yield at 
rapidities $|y| \geq 2$ is produced by fragmentation of the unscattered beam 
remnants, the model predicts a significant contribution to particle 
production at central rapidity from partons that have undergone perturbative 
scattering.\cite{Muller:GM97}  This contribution is rising rapidly with nuclear 
mass $A$, roughly as $(A_1A_2)^{1/3}$.  This perturbative contribution to 
the energy deposition at $|y| \leq 1$ coincides with a rapid increase of 
the energy density in scattered partons at $\tau <$ 1 fm/c, which rises from 
about 2 GeV/fm$^3$ in S+S to 5 GeV/fm$^3$ in Pb + Pb.  This rise may be 
correlated with the much enhanced suppression of charmonium production 
in Pb + Pb collisions as observed by the NA50 experiment.\cite{Muller:NA50}

\section{Open problems}

In spite of the considerable conceptual advances in the parton cascade
model compared to its original formulation, certain aspects have still
not been completely clarified. For issues concerning the initial state
I can only refer to the recent review article of McLerran and 
Venugopalan.\cite{Muller:ML99} One problem associated with the final
state concerns the source of the produced entropy: The classical gluon 
field radiated in the encounter of two specified color charge distributions 
retains its full coherence during the course of the 
interaction.\cite{Muller:MMR97}  The Gaussian
average over inital conditions formally introduces entropy at late
times, but a significant part of this entropy is already present in
the initial state. The problem of thermalization remains a theoretical
challenge even in this framework. It is likely that its solution must
be sought in the chaotic dynamical aspects of classical Yang-Mills
fields that have been observed in numerical simulations of the
classical non-abelian gauge theory.\cite{Muller:MT92,Muller:BMM95}
The apparent decoherence of classical gauge fields is visible in
a dramatic fashion when the collision of two non-abelian wave packets
is studied in numerical simulations on the lattice.\cite{Muller:PM99}

A consistent implementation of the RLSM into the parton cascade model
requires a transport description of partons including mean color fields.
Such a framework has been known for a long time 
\cite{Muller:Heinz85,Muller:EH89}, but practical implementations of
the QCD Boltzmann equation for partons and fields have been attempted 
only recently.\cite{Muller:MHM98}  The idea behind this approach is
to separate short--distance and long--distance dynamics by means of
a lattice cut-off: excitations with momenta $k \leq \pi/a$, where $a$ 
is the lattice spacing, are represented as classical fields on the
lattice; those with higher momenta are represented as colored particles.
The lattice cut-off must then be chosen such that $g\mu < \pi/a < \mu$.
``Hard'' collisions, i.e. those with a momentum transfer $q > \pi/a$,
are now infrared safe due to the lattice cut-off, soft collisions are
represented as interactions among particles via the lattice fields.
These incorporate the crucial screening effects at the scale $g\mu$.

Another conceptual issue that is understood, in principle, but whose
detailed investigation is an important outstanding problem, is the
question what constitutes a consistent set of semiclassical transport
equations for a theory with perturbatively massless modes such as QCD.
This issue is conceptually resolved in the case of a massless scalar
quantum field theory, where the introduction of a self-consistent 
medium--dependent mass term is sufficient. The issue is trickier in
the case of QCD because of two problems: the need to retain gauge
invariance and the long radiative tails of the spectral functions for
the colored quasi--particles. Gauge invariance essentially requires
that every modification of a $n$--body correlator is mirrored by an
analougous modification of the $(n+1)$--body correlator satisfing
Ward idenities. The treatment of radiative tails of the spectral
function, on the other hand, requires renormalization group techniques.

At the time of his death, Klaus was working on this problem, applying
the ``exact'' renormalization group technique \cite{Muller:Wet96} 
to QCD transport theory. On the evening before he
boarded the ill-fated Swissair flight, he sent an almost finished draft
of a manuscript \cite{Muller:Gei99} to Wetterich and me with the request: 
``I expect to see your comments when I return.'' I believe that, again,
Klaus was onto something very important, but it will remain for others
to finish what he started. Let me just end this part of the discussion
with the remark that the usual mass-shell condition in transport theory
is intimately connected with the renormalization--group equations, and 
that a better understanding of this connection is crucial to further
progress in the quantum transport theory of QCD.

The next logical step in any parton cascade model that aims at practical
predictions for experimentally observables is hadronization. In 
collaboration with John Ellis, Klaus worked intensely on this difficult 
problem.\cite{Muller:EG95}  Their work is based on an effective theory
of scale invariance breaking in QCD and has had some success in describing
hadronization of quark jets produced in $e^+e^-$ annihilations. Further
progress here will have to rely on an improved understanding of the 
mechanism of quark confinement in QCD.

Finally, let me point out that Steffen Bass had been in the midst of a
collaboration with Klaus to add a state-of-the-art hadronic cascade model 
(UrQMD \cite{Muller:Bass98}) to the parton cascade code VNI, in order to 
describe final--state interactions of the hadrons created by hadronization of
the quark--gluon plasma. The first results of this project are intriguing:
they indicate that a small fraction of the created hadrons emerge without
rescattering, whereas the bulk of the hadron yield is reprocessed through
the melting pot of hadronic reactions.\cite{Muller:Bass99}

\section{Summary}

The parton cascade model was developed by Klaus to provide a QCD-based 
description of the approach to a locally thermalized state in collisions 
of heavy ions in the RHIC energy regime and beyond.  In its original 
formulation the PCM predictions were critically dependent on several 
cut-off parameters that had to be determined from $pp$ collision data.  
Recent advances incorporating in--medium effects 
into the parton interactions have reduced this 
dependence significantly, possibly allowing the application of the PCM over 
a wider energy range.  Results obtained for nuclear collisions at CERN--SPS 
energies are intriguing. 

The in--medium effects that modify parton--parton interactions not only reduce 
the parameter dependence of the model, they also provide valuable insight 
into the dynamics of a dense parton plasma.  It is clear that we are just at 
the beginning here. The transport properties of off-shell quanta need to 
be understood much better, not only in cases where the off-shell propagator 
is dominated by a well-defined resonance, but especially in the case where 
the particles never get close to their mass shell as it applies to QCD.  
Another open question concerns the need for mean color fields.  Such fields 
are not included in present versions of the PCM, but the random light--cone 
source model suggest that mean fields may be essential ingredients of a 
complete description of soft processes in nuclear collisions.  One needs 
to take an average over a Gaussian ensemble of mean fields, where the width 
of the field distribution is more important than the expectation value 
which remains zero.  It would be interesting to explore possible connections 
of the RLSM to the traditional chromo--hydrodynamical 
model.\cite{Muller:BC85,Muller:GKM85} 

Ultimately, the question is whether the parton cascade model can be replaced 
by a controlled approximation scheme where, in principle, successive orders 
of ever more sophisticated corrections can be calculated.  We are still some 
steps away from a consistent formulation of transport phenomena off equilibrium 
in QCD. It is even unclear whether we even know what the small parameters 
in such an approximation scheme are.  It is clear that a high 
density of excitations of the QCD vacuum is an essential condition, but there 
are many subtleties if one wants to go beyond this statement.  However, steady 
progress in this field is being made, and there is reason to hope that a 
consistent formulation of transport phenomena in QCD can ultimately be 
achieved. 

Finally, the treatment of the late phase of a relativistic heavy ion 
collision, when the dense matter breaks up into individual hadrons,
has recently seen some exciting improvements.  One still needs to rely
on a phenomenological hadronization model for the transition to a 
hadronic cascade, but the latter can be studied with he technology
that has been developed for the description of nuclear collisions at
lower energies.  In fact, the description of the hadronic final--state
interactions should be much more reliable than, e.g.~that of a nuclear
collision at AGS energies, because the hadronic system starts out close
to thermal equilibrium.  Klaus was very excited about this approach,
which will allow for a better comparison of the predictions of the parton 
cascade model with the experimental data that will soon become available
from RHIC.

\section*{Acknowledgments}
This work was supported in part by grant no. DE-FG02-96ER40945 from the 
U.S.\ Department of Energy.


\begin{thebibliography}{99} 

\bibitem{Muller:GM92}  
K. Geiger and B. M\"uller, {\em Nucl. Phys.} {\bf B369}, 600 (1992)

\bibitem{Muller:Gei92a}  
K. Geiger, {\em Phys. Rev.} {\bf D46}, 4965 and 4986 (1992)

\bibitem{Muller:Gei92b}  
K. Geiger, {\em Phys. Rev.} {\bf D47}, 133 (1993)

\bibitem{Muller:Gei94} 
K. Geiger, B. M\"uller, {\em Phys. Rev.} {\bf D50}, 357 (1994); K. Geiger, 
{\em Phys. Rev.} {\bf D50}, 3243 (1994)

\bibitem{Muller:Gei95} 
K. Geiger, {\em Phys. Rep.} {\bf 258}, 378 (1995) 

\bibitem{Muller:AKM80} 
R. Anishetty, P. Koehler, and L. McLerran, {\em Phys. Rev.} {\bf D22},
 2793 (1980)

\bibitem{Muller:Bj83}  
J.D. Bjorken, {\em Phys. Rev.} {\bf D27}, 140 (1983)

\bibitem{Muller:HK86} 
R.C. Hwa and K. Kajantie, {\em Phys. Rev. Lett.} {\bf 56}, 696 (1986) 

\bibitem{Muller:BM87}  
J.P. Blaizot and A.H. Mueller, {\em Nucl. Phys.}, {\bf B289}, 847 (1987) 

\bibitem{Muller:KLL87} 
K. Kajantie, P.V. Landshoff, and J. Lindfors, {\em Phys. Rev. Lett.} {\bf 59}
, 2527 (1987)

\bibitem{Muller:EKL} 
K.J. Eskola, K. Kajantie, and J. Lindfors, {\em Phys. Lett.} {\bf B214}, 613 
(1988); {\em Nucl. Phys.} {\bf B323}, 37 (1989)

\bibitem{Muller:GW}
M. Gyulassy and X.N. Wang, {\em Phys. Rev. } {\bf D44}, 3501 (1991)

\bibitem{Muller:Wang97} 
X.N. Wang, {\em Phys. Rep.} {\bf 280}, 287 (1997)

\bibitem{Muller:Gei96}  
K. Geiger, {\em Phys. Rev.} {\bf D56}, 2665 (1997)

\bibitem{Muller:XS94}   
L. Xiong  and E.V. Shuryak, {\em Phys. Rev. } {\bf C49}, 2203 (1994)

\bibitem{Muller:EW94}
K.J. Eskola and X.N. Wang, {\em Phys. Rev.} {\bf D49}, 1284 (1994) 

\bibitem{Muller:HW96}
H. Heiselberg and X.N. Wang, {\em Nucl. Phys.} {\bf B462}, 389 (1996); 
{\em Phys. Rev.} {\bf C53}, 1892 (1996)

\bibitem{Muller:MR96} 
J. Rau and B. M\"uller, {\em Phys. Rep.} {\bf 272}, 1 (1996)

\bibitem{Muller:Baier}  
R. Baier, Yu. L. Dokshitzer, S. Peign\'e, and D. Schiff, {\em Phys. Lett.} 
{\bf B345} 277 (1995); 
R. Baier, Yu. L. Dokshitzer, A.H. Mueller, S. Peign\'e and D. Schiff, 
{\em Nucl. Phys.} {\bf B483}, 291 (1997);  {\bf B484}, 265 (1997)

\bibitem{Muller:MV94}  
L. McLerran and R. Venugopalan, {\em Phys. Rev.} {\bf D49}, 2233 and 3352 
(1994)

\bibitem{Muller:Kov96}  
Yu.V. Kovchegov, {\em Phys. Rev.} {\bf D54}, 5463 (1996); {\bf D55}, 5455 
(1997) 

\bibitem{Muller:KR96}   
Yu.V. Kovchegov and D.H. Rischke, {\em Phys. Rev.} {\bf C56}, 1084 (1997)

\bibitem{Muller:MMR97}  
S.G. Matinyan, B. M\"uller and D.H. Rischke, {\em Phys. Rev.} {\bf C56}, 2191 
(1997); {\bf C57}, 1927 (1998)

\bibitem{Muller:GML97}  
M. Gyulassy and L. McLerran, {\em Phys. Rev.} {\bf C56}, 2219 (1997)

\bibitem{Muller:KMW95}  
A. Kovner, L. McLerran, and H. Weigert, {\em Phys. Rev.} {\bf D52}, 3809 and  
6231 (1995)

\bibitem{Muller:KMW97}
J. Jalilian-Marian, A. Kovner, L. McLerran, and H. Weigert, {\em Phys. Rev.} 
{\bf D55}, 5414 (1997) 

\bibitem{Muller:KLW97}  
J. Jalilian-Marian,   A. Kovner, A. Leonidov, and H. Weigert, {\em Nucl. 
Phys. }  {\bf B504},  415 (1997)

\bibitem{Muller:Kov98}
Yu.V. Kovchegov and A.H. Mueller, {\em Nucl. Phys. B\bf 529}, 451 (1998)
 
\bibitem{Muller:Qiu}
J.W. Qiu, {\em Nucl. Phys.} {\bf B291}, 746 (1987) 

\bibitem{Muller:EQW} 
K.J. Eskola, J. Qiu, and X.N. Wang, {\em Phys. Rev. Lett.} {\bf 72}, 36 (1994) 

\bibitem{Muller:BMW92} 
T.S. Bir\'{o}, B. M\"uller, and X.N. Wang, {\em Phys. Lett.} {\bf B283}, 171 
(1992)

\bibitem{Muller:Zak}  
B.G. Zakharov, JETP Letters {\bf 63}, 952 (1996); {\bf 64}, 781 (1996); 
{\bf 65}, 615 (1997)

\bibitem{Muller:KHN} 
J. H\"ufner, B. Kopeliovich and J. Nemchik, {\em Phys. Lett.} 
{\bf B383}, 362 (1996)

\bibitem{Muller:EMW96}  K.J. Eskola, B. M\"uller, and X.N. Wang, {\em Phys. Lett.} 
{\bf B374}, 20 (1996)

\bibitem{Muller:VNI}
K. Geiger, {\em Comp. Phys. Comm.} {\bf 104}, 70 (1997) 
 
\bibitem{Muller:BD98} 
D.A. Brown and P. Danielewicz, preprint nucl-th/9802015 

\bibitem{Muller:MS98} 
A. Makhlin and E. Surdutovich, preprint hep-ph/9803364 

\bibitem{Muller:Biro93}
T.S. Bir\'o, E. van Doorn, B. M\"uller, M.H. Thoma, and X.N. Wang, 
{\em Phys. Rev.} {\bf C48}, 1275 (1993) 

\bibitem{Muller:Stri94}
M.T. Strickland, {\em Phys. Lett.} {\bf B331}, 245 (1994)

\bibitem{Muller:SMM97} 
D.K. Srivastava, M.G. Mustafa, and B. M\"uller, {\em Phys. Rev.} {\bf C56}, 
1064 (1997)

\bibitem{Muller:GS97}  
K. Geiger and D.K. Srivastava, {\em Phys. Rev.} {\bf C56}, 2718 (1997) 

\bibitem{Muller:GM97}  
K. Geiger and B. M\"uller, {\em Heavy Ion Physics } {\bf 7}, 207 (1998) 

\bibitem{Muller:NA50} 
M. Gonin, (NA50 collaboration), {\em Nucl. Phys.} {\bf A610}, 404c (1996)

\bibitem{Muller:BC85} 
A. Bialas and W. Czyz, {\em Phys. Rev.} {\bf D30}, 2371 (1984); {\bf D31}, 
198 (1985)

\bibitem{Muller:GKM85}  
G. Gatoff, A.K. Kerman, and T. Matsui, {\em Phys. Rev.} {\bf D36}, 
114 (1987)

\bibitem{Muller:ML99}
L. McLerran and R. Venugopalan, preprint hep-ph/9809427

\bibitem{Muller:MT92}
B. M\"uller and A. Trayanov, {\em Phys. Rev. Lett.} {\bf 68}, 3387 (1992) 

\bibitem{Muller:BMM95}
T.S. Bir\'o, S.G. Matinyan, and B. M\"uller, {\em Chaos in Gauge Field
Theory} (World Scientific, Singapore, 1995)

\bibitem{Muller:PM99}
W. P\"oschl and B. M\"uller, preprint nucl-th/9812066

\bibitem{Muller:Heinz85}
U. Heinz, {\em Ann. Phys.} (NY) {\bf 161}, 48 (1985); {\bf 168}, 148 (1986)

\bibitem{Muller:EH89}
H.T. Elze and U. Heinz, {\em Phys. Rep.} {\bf 183}, 81 (1989) 

\bibitem{Muller:MHM98}
G.D. Moore, C.R. Hu, and B. M\"uller, {\em Phys. Rev.} {\bf D58}, 045001
(1998)

\bibitem{Muller:Wet96}
C. Wetterich, {\em Phys. Lett. B}{\bf 301}, 90 (1993)

\bibitem{Muller:Gei99}
K. Geiger, preprint hep-ph/9902289

\bibitem{Muller:EG95}
J. Ellis and K. Geiger, {\em Phys. Rev.} {\bf D52}, 1500 (1995)

\bibitem{Muller:Bass98}
S.A. Bass, et al., {\em Prog. Part. Nucl. Phys.} {\bf 41}, 255 (1998)

\bibitem{Muller:Bass99}
S.A. Bass, M. Hofmann, M. Bleicher, L. Bravina, E. Zabrodin, H. St\"ocker, 
and W. Greiner, preprint nucl-th/9902055

\end{thebibliography}
\end{document}